\documentclass[12pt]{iopart}

\usepackage{graphicx}
\usepackage{iopams}  

\usepackage{color}

\usepackage[normalem]{ulem}
\usepackage{cancel}

\begin{document}

\title[Fluctuation relation in continuous-time random walks]{Fluctuation relation in continuous-time random walks driven by an external field}

\author{Kazuhiko Seki}

\address{National Institute of Advanced Industrial Science and Technology (AIST), Onogawa 16-1 AIST West, Ibaraki 305-8569, Japan
}
\ead{k-seki@aist.go.jp}
\vspace{10pt}
\begin{indented}
\item[]August 2023
\end{indented}

\begin{abstract}
We study a fluctuation relation representing a nonequilibrium equality indicating that the ratio between the distribution of trajectories obtained by exchanging the initial and final positions is characterized by free energy differences for the duration of the trajectories.
We examine the fluctuation relation for noninteracting charge carriers driven by an external electric field by using 
a continuous-time lattice random walk model with a general waiting-time distribution of transitions. The fluctuation relation is obtained regardless of the lattice structure factor or the form of the waiting-time distribution. 
However, the fluctuation relation is satisfied only after taking the continuum limit in the presence of a reflecting boundary. Moreover, 
in free space without boundary conditions, exchanging the initial and final positions is equivalent to exchanging the field (or drift) directions. However, we show that the exchanging field (or drift) directions is not relevant for studying the fluctuation relation under the reflecting boundary condition.
\end{abstract}

%
%
%
%
%
\section{Introduction}

Over the past several decades, fluctuation theorems that enhance our understanding of the relation between nonequilibrium thermodynamics and statistics in the fluctuation of trajectories have been formulated. \cite{Bochkov_77,BOCHKOV_81,Evans_93,Evans_94,Evans_02,Gallavotti_95,Kurchan_98,Searles_99,Crooks_98,Crooks_99,Crooks_00,Jarzynski_PRE,Seifert_05,Seifert_12,Zon_03,Horowitz_07,Seifert_08}
The proposed fluctuation theorems require neither the adiabatic operation commonly imposed for equilibrium thermodynamic relations nor a limited strength of the applied field, as imposed for the linear response theorem. 
The early fluctuation theorem was formulated in terms of entropy production: the probability of observing trajectories opposing positive entropy production decreases exponentially. \cite{Evans_93,Evans_94,Evans_02}
Later, the Crooks fluctuation theorem was formulated in terms of the free energy difference: 
the probability of observing trajectories opposing a positive free energy difference decreases exponentially. \cite{Crooks_98,Crooks_99,Crooks_00}
The Jarzynski equality relating the free energy difference and the work done \cite{Jarzynski_PRE} enables the Crooks fluctuation theorem to be interpreted as indicating that the ratio between the distribution of trajectories obtained by exchanging the initial and final states is characterized by  
the work done for the duration of the trajectories. \cite{Crooks_98,Crooks_99,Crooks_00} 

Fluctuation theorems have been proposed through studies of the phase-space Jacobian volume element of trajectories using the Liouville equation for deterministic systems, numerical calculations for many-particle systems, and the Langevin equation for stochastic systems, where particles move continuously in space. \cite{Evans_93,Evans_94,Evans_02,Gallavotti_95,Kurchan_98,Searles_99,Crooks_98,Crooks_99,Crooks_00,Jarzynski_PRE}
Fluctuation theorems have also been studied using master equations for discrete systems by assuming Markovian time evolution. \cite{Jarzynski_PRE,Esposito_06}
In some cases, the path integral formulation has been used to propose fluctuation theorems by assuming Markovian time evolution. \cite{Chernyak_06,Taniguchi_07}
The fluctuation theorem has also been studied using the generalized Langevin equation or Fokker--Planck equation, where non-Markovian stochastic dynamics is assumed. \cite{Ohkuma_07,Speck_07,Mai_07,Chaudhury_08,Chechkin_09,Dieterich_15,Aquino_15}
The path integral formulation has been also developed to study the fluctuation theorem using generalized Langevin equation by evaluating the Jacobian arising from non-Markovian response. \cite{Ohkuma_07} 

In continuous time random walks,   
two conditions are required for microscopic reversibility and the fluctuation theorem; 
one is independence of transition direction and waiting time for transition rates (separability of the waiting time distribution) and the other condition is detailed balance. 
\cite{Qian_06,Wang_07,Esposito_08}
We consider the cases that both conditions are satisfied and focus on the effect of a reflecting boundary on the fluctuation theorem. 
Previously, the fluctuation relation has been presented by taking into account the lattice structure and a memory kernel 
in continuous time lattice random walks. \cite{Berezhkovskii_08,burov_22} 
The results are further extended to study the effects of branched states and the channels on the fluctuation theorem. \cite{Hamid_13}
A thermodynamic interpretation to a fluctuation theorem derived from continuous time random walks was also given. \cite{Esposito_08}

However, the coupled effects of the lattice structure factor for a random walk, a memory kernel, and the boundary conditions have not yet been fully examined. 
Here, we use continuous-time lattice random walk models with a general waiting-time distribution to study the random walk of noninteracting charge carriers driven by an external electric field.
We show a fluctuation relation by expressing transition probabilities for the continuous-time random walk models in a suitable form to study the effect of exchanging the initial and final positions.
The explicit expressions for transition probabilities enable us to clarify the difference between exchanging initial and final positions and exchanging field (or drift) directions.
We show that the ratio between the usual transition probability and the probability obtained by exchanging the initial and final positions is related to the free energy difference between the initial and final positions, irrespective of the lattice structure factor and the form of the waiting time distribution. 
The results suggest that the Crooks fluctuation theorem 
 holds for continuous-time lattice random walks 
in free space without boundaries, as demonstrated by various methods. \cite{Esposito_08,Berezhkovskii_08,burov_22,Hamid_13} However, in the presence of a reflecting boundary, the Crooks fluctuation theorem does not hold for continuous time random walk models if a continuum limit is not taken. The results suggest that the separability of the waiting time distribution and the detailed balance condition are sufficient conditions for the Crooks fluctuation theorem to hold in a continuous time random walk, as long as there is no reflecting boundary. Another aspect of introducing a reflecting boundary is also obtained. 
In free space without boundary conditions, exchanging the initial and final positions is equivalent to exchanging the field (or drift) directions. However, we show that exchanging the field (or drift) directions is not relevant for studying the fluctuation relation under a reflecting boundary condition.



\section{Drift diffusion in free space}
\begin{figure}[h]
\begin{center}
\includegraphics[width=10cm]{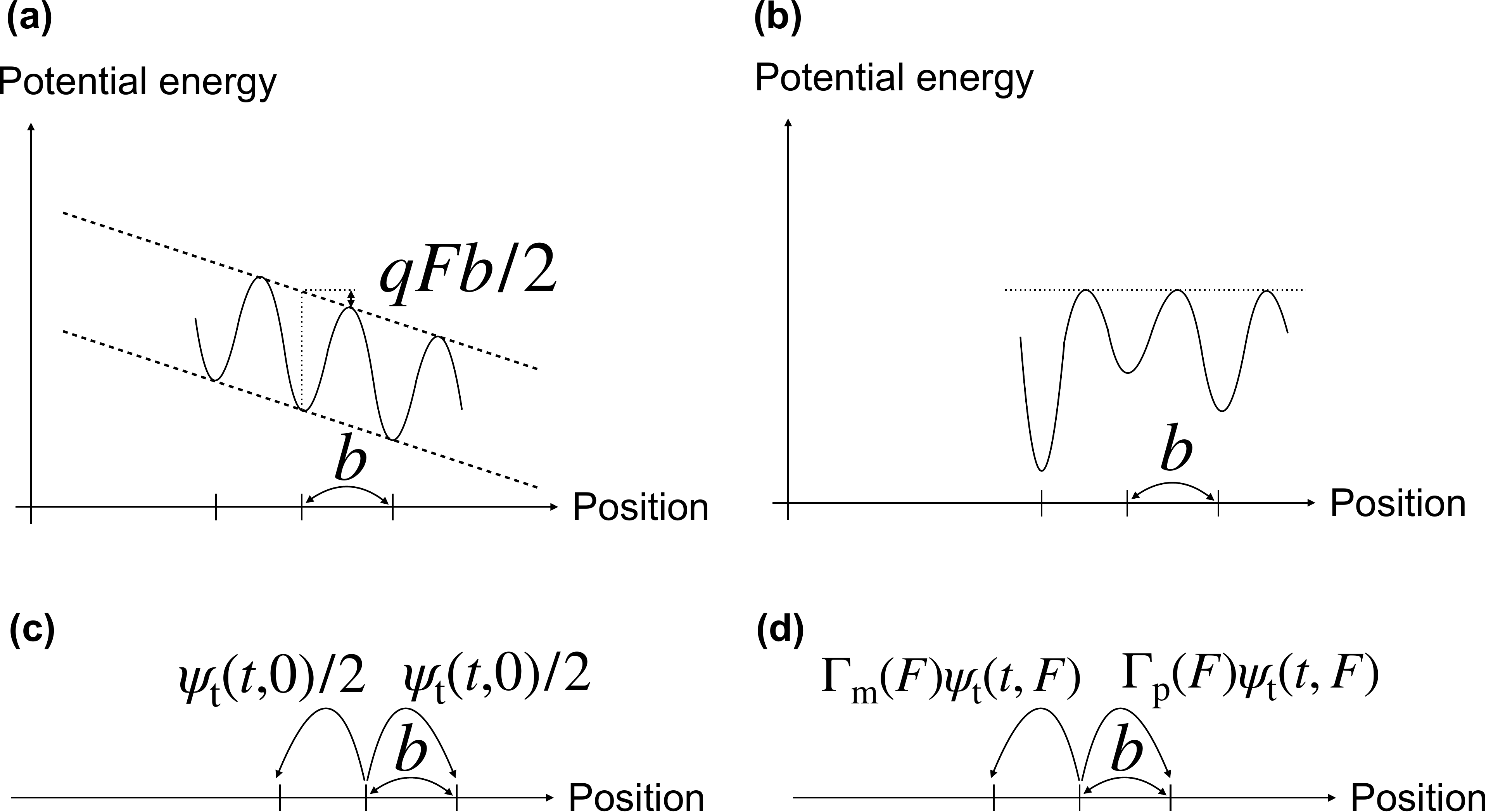}
\end{center}
\caption{(a) and (b) Schematics of the potential energy for transitions under an external electric field ($F$) applied in the direction of increasing $x$; $b$ indicates the lattice constant. 
(c) and (d) Continuous-time random walk model. 
(a) Energy landscape without energetic disorder. 
The activation energy is lowered by $q F b/2$ in the transition-rate 
($\gamma_{\rm rp}(F)$) along the direction of the external field, and raised by $q F b/2$ in the transition-rate 
($\gamma_{\rm rm}(F)$) along the opposite direction.
The local detailed balance condition is satisfied, {\it i.e.}, $\gamma_{\rm rp}(F)/\gamma_{\rm rm}(F)=\exp\left[ q Fb/(k_{\rm B} T)\right] $. 
(b) Energy landscape under energetic disorder when the applied external field is absent. 
(c) The effect of energetic disorder can be taken into account by the total waiting-time distribution denoted by $\psi_{\rm t} (t,F=0)$ when $F=0$. 
(d) The superposition of the potential energy landscape of (a) and that of (b) can be taken into account using $\psi_{\rm t} (t,F)$:
$\Gamma_{\rm p} (F) = \gamma_{\rm rp}(F)/\gamma_{\rm rt}(F)$ and $\Gamma_{\rm m} (F)= \gamma_{\rm rm}(F)/\gamma_{\rm rt}(F)$,
where $\gamma_{\rm rt}(F)=\gamma_{\rm rp} (F)+\gamma_{\rm rm} (F)$. 
}
\label{fig:1}
\end{figure}

We consider the fluctuation relation in a one-dimensional random walk along the $x$-axis under energetic disorder and an external electric field with strength $F$.
Here, we study nonequilibrium states in which charge carriers flow without interaction among themselves under the external electric field and without any boundary.
When energetic disorder presents, the waiting-time distribution might have a power-law tail; we study the effect of a heavy-tail waiting-time distribution of transitions on the fluctuation theorem.
We also study the effect of a reflecting boundary condition and the effect of the structure factor for a random walk on the fluctuation theorem.
We take the direction of the field as the $x$-axis and denote $F$ as the electric field strength acting on the charge carriers; the charge on each carrier is denoted by $q$. 
First, we focus on the field dependence alone without energetic disorder in one-dimensional transitions [Fig. \ref{fig:1}(a)]. 
For convenience,  we introduce $\gamma_{\rm r}$ when an applied field is absent;  
$\gamma_{\rm r}$ indicates 
the transition rate 
from a trap to both neighboring sites. 
Under an applied field, 
the transition-rate 
in the field direction and 
that in the opposite direction are denoted by $\gamma_{\rm rp}(F)$ and $\gamma_{\rm rm}(F)$, respectively. 
By considering the Arrhenius law, these two transition-rates 
can be expressed as \cite{Seki_23}
\begin{eqnarray}
\gamma_{\rm rp}(F)&=(\gamma_{\rm r} /2)\exp\left[ q Fb/(2k_{\rm B} T)\right] 
\label{eq:gammap}
\\
\gamma_{\rm rm}(F)&=(\gamma_{\rm r}/2) \exp\left[ -q Fb/(2k_{\rm B} T)\right] ,
 \label{eq:gammam}  
 \end{eqnarray}
where $q$, $b$, $k_{\rm B}$ and $T$ are the elementary charge, the lattice constant, the Boltzmann constant and temperature, respectively. 
$\gamma_{\rm rp}(F)$ and $\gamma_{\rm rm}(F)$ satisfy the local detailed balance condition: {\it i.e.}, $\gamma_{\rm rp}(F)/\gamma_{\rm rm}(F)=\exp\left[ q Fb/(k_{\rm B} T)\right] $.
The factor two in $(\gamma_{\rm r}/2)$ is required because $\gamma_{\rm r}$ represents transition from a trap to both neighboring sites. 
Under the influence of an external electric field, the expression for the total transition frequency ($\gamma_{\rm r}$) changes to 
\begin{eqnarray}
\gamma_{\rm rt}(F)=\gamma_{\rm rp} (F)+\gamma_{\rm rm} (F)=\gamma_{\rm r}  \cosh\left[ q Fb/(2k_{\rm B} T)\right] 
\label{eq:gamma_rt},
\end{eqnarray}
which reduces to $\gamma_r$ when $F=0$.

Next, we consider that carriers undergo transitions from the state with random trapping energy $E$. 
Without an external field [Fig. \ref{fig:1}(b)], the release rate is proportional to $\exp \left[ - E/(k_{\rm B} T) \right]$, where the distribution function for the trap energy is denoted by $g(E)$.
Under an external field, the detrapping rate in the field direction can be expressed as $\gamma_{\rm p} (E,F)= \gamma_{\rm rp} (F)\exp \left[ - E/(k_{\rm B} T) \right] $.
In addition, the detrapping rate in the opposite direction can be expressed as $\gamma_{\rm m} (E,F)= \gamma_{\rm rm} (F)\exp \left[ - E/(k_{\rm B} T) \right] $.
The total detrapping rate can be defined by 
\begin{eqnarray}
\gamma_{\rm t} (E,F) =
\left[ \gamma_{\rm rp} (F)+\gamma_{\rm rm} (F) \right]\exp \left[ - E/(k_{\rm B} T) \right] .
\label{releasert}
\end{eqnarray}

In the continuous-time random walk (CTRW) model, the energetic disorder is taken into account by the waiting-time distribution, as shown in Fig. \ref{fig:1}(c) for $F=0$. 
The waiting time distribution in the continuous-time random walk can be formulated by 
assuming 
 annealed energy disorder, where 
trap energy distribution [$g(E)$] is used in every transition step. 
(If each site has the fixed random trap energy, the situation is called quenched energy disorder, where the particle may return to the previously occupied site and 
the trap energy needs not necessarily be renewed. )
Under both an external field and energetic disorder [Fig. \ref{fig:1}(d)], the total waiting-time distribution can be defined as
\begin{eqnarray}
\psi_{\rm t} (t,F) = 
\int_0^{\infty} d\,E g(E) \gamma_{\rm t} (E,F)  \exp \left( - \gamma_{\rm t} (E,F)  t \right) .
\label{eq:psitt}
\end{eqnarray}
The waiting-time distribution along the field direction is expressed by 
$\psi_{\rm p} (t,F) =\Gamma_{\rm p} (F) \psi_{\rm t} (t,F) $, where $\Gamma_{\rm p} (F)$ indicates the fraction of transitions from $x$ to $x+b$ against the sum of the transitions to both neighboring sites: 
\begin{eqnarray}
\Gamma_{\rm p} (F) &= \frac{\gamma_{\rm rp}(F)}{\gamma_{\rm rt}(F)} =\frac{\exp\left[ q Fb/(2k_{\rm B} T)\right]}{2\cosh\left[ q Fb/(2k_{\rm B} T)\right]}. 
\label{eq:psip}
\end{eqnarray}
Similarly, the waiting-time distribution along the opposite direction is expressed by $\psi_{\rm m} (t,F) =\Gamma_{\rm m} (F) \psi_{\rm t} (t,F) $, 
where $\Gamma_{\rm m} (F)$ indicates 
\begin{eqnarray}
\Gamma_{\rm m} (F)&=  \frac{\gamma_{\rm rm}(F)}{\gamma_{\rm rt}(F)} =\frac{\exp\left[ -q Fb/(2k_{\rm B} T)\right]}{2\cosh\left[ q Fb/(2k_{\rm B} T)\right]}.
\label{eq:psim}
\end{eqnarray}
Using Eq. (\ref{eq:psip}) and (\ref{eq:psim}), we find the local detailed balance,   
\begin{eqnarray}
\frac{\Gamma_{\rm p} (F) }{\Gamma_{\rm m} (F) }=
\exp\left[ q Fb/(k_{\rm B} T)\right].
\label{eq:15_5_st}
\end{eqnarray}

For the biased CTRW model in free space without any boundary, the Green's function with the initial position denoted by $x_{\rm i}$ should be translationally invariant and 
$G_0(x,x_{\rm i},t)$ in the Laplace domain can be expressed as \cite{Montroll,Hughes_95}
\begin{eqnarray}
\hat{G}_0(x,x_{\rm i},s)&= \frac{1-\hat{\psi}_{\rm t}}{s}\frac{1}{2\pi} \int_{-\pi/b}^{\pi/b} dk\, \frac{\exp[-ik(x-x_{\rm i})]}{1-\hat{\psi}_{\rm t} \lambda(k)},
\label{eq:G0_0d}
\end{eqnarray}
where the Laplace transform of time-dependent function $f(t)$ is denoted by $\hat{f}(s)$, $\hat{\psi}_{\rm t}$ is the Laplace transform of $ \psi_{\rm t} (t,F)$, and $\lambda(k)$ is the structure factor for a biased random walk in one-dimension,
\begin{eqnarray}
\lambda(k)=\Gamma_{\rm p} (F) \exp\left(i k b \right) +\Gamma_{\rm m} (F)\exp\left(-i k b \right) .
\label{eq:lambda}
\end{eqnarray}
$G_0(x,x_{\rm i},t)$ can be expressed using the lattice Green's function with modification by the waiting-time distribution to express the probability of arriving at $x$ before $t$ starting from $x_{\rm i}$. 
The lattice Green's function modified by the waiting-time distribution is given by \cite{Hughes_95}
\begin{eqnarray}
\hat{G}_{L0}(x,x_{\rm i},s)&= \frac{1}{2\pi} \int_{-\pi/b}^{\pi/b} dk\, \frac{\exp[-ik(x-x_{\rm i})]}{1-\hat{\psi}_{\rm t} \lambda(k)} 
\nonumber \\
&= \frac{1}{2\pi} \int_{-\pi/b}^{\pi/b} dk\, \exp[-ik(x-x_{\rm i})]\sum_{j=0}^\infty \left[\hat{\psi}_{\rm t} (s) \lambda(k) \right]^j . 
\label{eq:G0_0dL}
\end{eqnarray}
$[\hat{\psi}_{\rm t} (s)]^j$ becomes the time convolution after the inverse Laplace transformation; the inverse Laplace transform of Eq. (\ref{eq:G0_0d}) can be interpreted as the sum of the transition probability corresponding to $j$-times transitions between $x_{\rm i}$ and $x$, which occurs until time $t$. 
In Eq. (\ref{eq:G0_0d}), the factor $(1-\hat{\psi}_{\rm t})/s$ is the Laplace transform of $\varphi(t,F)=\int_t^\infty dt_1 \psi_{\rm t}(t_1,F)$, which indicates the remaining probability of a carrier not transitioning to a new site until $t$. 
The usual Green's function given by Eq. (\ref{eq:G0_0d}) is obtained from $G_0(x,x_{\rm i},t)=\int_0^t dt_1 \varphi(t-t_1)G_{L0}(x,x_{\rm i},t_1)$; the lattice Green's function is convoluted with the remaining probability of the carrier arriving at $x$ and remaining for time $t$. 

For an unbiased random walk in one-dimension, we obtain $\lambda (k)=(1/2)\sum_{x=\pm b} \exp(i k x)\approx 1-(kb)^2/2$, where we assumed $kb\ll1$. 
Under the bias, 
we can express the structure factor $\lambda (k)$ by assuming $kb\ll1$,  
\begin{eqnarray}
\lambda(k) \approx 1 +i\mu k-\frac{\sigma^2}{2} k^2, 
\label{eq:C}
\end{eqnarray}  
where $\mu$ is related to the drift and $\sigma^2  $ is related to the dispersion; they are given by
\begin{eqnarray}
\mu&=b\left[\Gamma_p (F)-\Gamma_m(F)\right]=b\tanh[qFb/(2k_{\rm B} T)]\approx qFb^2/(2k_{\rm B} T),
\label{eq:Aapprox}\\
\sigma^2  &=b^2 .
\label{eq:Bapprox}
\end{eqnarray}
Under the assumption $kb\ll1$, $\hat{G}_0(x,x_{\rm i},s)$ can be approximated as \cite{Seki_23}
\begin{eqnarray}
\hat{G}_0(x,x_{\rm i},s)&\approx \frac{1-\hat{\psi}_{\rm t}}{s}\frac{1}{2\pi} \int_{-\infty}^\infty dk\, \frac{\exp[-ik(x-x_{\rm i})]}{1-\hat{\psi}_{\rm t} \left[1+i\mu k-(\sigma^2  /2)k^2 \right]}.
\label{eq:G0}
\end{eqnarray}

We now transform $\hat{G}_0(x,x_{\rm i},s)$ into a form suitable to study the fluctuation relation.
By introducing $k_1=k-i\mu/\sigma^2  $, we obtain 
\begin{eqnarray}
-i\mu k+(\sigma^2  /2)k^2 =\mu^2/(2\sigma^2  )+(\sigma^2  /2) k_1^2, 
\label{eq:abs_5}
\end{eqnarray}
and Eq. (\ref{eq:G0}) can be rewritten as
\begin{eqnarray}
\hat{G}_0(x,x_{\rm i},s)&= \frac{1-\hat{\psi}_{\rm t}}{s}\frac{1}{2\pi} \int_{-\infty}^\infty dk_1\, \frac{\exp[-i(k_1+i\mu /\sigma^2  )(x-x_{\rm i})]}{1-\hat{\psi}_{\rm t} \left[1+\mu^2/(2\sigma^2  )+(\sigma^2  /2) k_1^2 \right]} .
\label{eq:abs_6_2}
\end{eqnarray}
Therefore, we obtain \cite{Seki_23}
\begin{eqnarray}
G_0(x,x_{\rm i},t)=\exp\left[(\mu /\sigma^2  )(x-x_{\rm i})\right]g_0(|x-x_{\rm i}|,t), 
\label{eq_7}
\end{eqnarray}
where $g_0(|x-x_{\rm i}|,t)$ is the inverse Laplace transform of $\hat{g}_0(|x-x_{\rm i}|,s)$,  \cite{Seki_23}
\begin{eqnarray}
\hat{g}_0(|x-x_{\rm i}|,s)&= \frac{1-\hat{\psi}_{\rm t}}{s}\frac{1}{2\pi} \int_{-\infty}^\infty dk\, \frac{\exp[-ik(x-x_{\rm i})]}{1-\hat{\psi}_{\rm t} \left[1+\mu^2/(2\sigma^2  )+(\sigma^2  /2) k^2 \right]} .
\label{eq:abs_8}
\end{eqnarray}
$\hat{g}_0(x-x_{\rm i},s)$ is an even function of $x-x_{\rm i}$.
Using the Green's function, we derive the fluctuation relation: \cite{Seki_23}
\begin{eqnarray}
\frac{G_0(x,x_{\rm i},t)}{G_0(x_{\rm i},x,t)}&=
\exp\left(2\tanh[qFb/(2k_{\rm B} T)](x-x_{\rm i})/b
\right)
\label{eq;fl0}\\
&\approx \exp \left(\frac{qF(x-x_{\rm i})}{k_{\rm B} T}\right) , 
\label{eq;fl}
\end{eqnarray}
where the initial condition is given by $x_{\rm i}$ at $t=0$. 
If we use Eq. (\ref{eq:Aapprox}) without the approximation given by $qFb/(2k_{\rm B} T)<1$, then $qF(x-x_{\rm i})/(k_{\rm B} T)$ on the right-hand side of Eq. (\ref{eq;fl}) is replaced with $2\tanh[qFb/(2k_{\rm B} T)](x-x_{\rm i})/b$ [Eq. (\ref{eq;fl0})]. 
As we will show later by Eq. (\ref{eq;fll}), Eq. (\ref{eq;fl}) rigorously holds if we retain the discrete nature of transitions even for $qFb/(2k_{\rm B} T)>1$. 
The results indicate that the limit of $qFb/(2k_{\rm B} T)<1$ is necessary if we approximate the structure factor using Eq. (\ref{eq:C}). 
That is, the continuum limit of $b \rightarrow 0$ should not be taken to study the nonlinear field dependence.

The relation given by Eq. (\ref{eq;fl}) represents the nonequilibrium equality characterized by free energy differences. \cite{Jarzynski_97}
$qF(x-x_{\rm i})/(k_{\rm B} T)$ in Eq. (\ref{eq;fl}) can be interpreted as the free energy difference between $x$ and $x_{\rm i}$, induced by the external electric field divided by $k_{\rm B} T$. 

Although the normalization constant for a steady-state distribution can be defined only formally in infinite systems, we study the ratio between the steady-state distribution at $x$ and that at $x_{\rm i}$ using the common normalization constant formally introduced; the normalization constant cancels out in taking the ratio. 

If the ratio between the stationary distribution at $x$ and that at $x_{\rm i}$ is replaced by the ratio of the equilibrium distribution at $x$ to that at $x_{\rm i}$ in a closed bounded system, 
then Eq. (\ref{eq;fl}) represents the detailed balance condition. \cite{vanKampen}
However, $G_0(x,x_{\rm i},t)$ depends on the boundary conditions. 
Here, we consider the open systems without boundary conditions and particles are driven by an external field. 
By considering that the free energy change on the right-hand side of Eq. (\ref{eq;fl}) can be rewritten as the ratio between the stationary distribution at $x$ and that at $x_{\rm i}$, we can regard Eq. (\ref{eq;fl}) as the extended detailed balance between the positions $x$ and $x_{\rm i}$.

We note that $G_0(x,x_{\rm i},t)$ [Eq. (\ref{eq_7})] is a function of $x-x_{\rm i}$, 
which allows us to express ${\cal G}_0(x-x_{\rm i},t)=G_0(x,x_{\rm i},t)$.
Equation (\ref{eq;fl}) can be rewritten as
\begin{eqnarray}
\frac{{\cal G}_0(y,t)}{{\cal G}_0(-y,t)}=\exp \left(\frac{qFy}{k_{\rm B} T}\right), 
\label{eq;fld}
\end{eqnarray}
where we defined $y=x-x_{\rm i}$. 
Equation (\ref{eq;fld}) has been regarded as a form of the fluctuation theorem. \cite{Chechkin_09,Berezhkovskii_08,burov_22,Hamid_13} 
It has been pointed out that Eq. (\ref{eq;fld}) is independent of the form of the waiting time distribution owing to the subordination principle;  \cite{Chechkin_09} 
the subordination relation holds for the present continuous time random walk models. \cite{Saichev_97,Barkai_01,Meerschaert_04,Sokolov_05,Yuste_05,Chechkin_09}
Equation (\ref{eq;fld}) allows the ratio of the transition probability for arriving at $y$ from the initial location to that for arriving at $-y$ from the initial location to be interpreted as being equal to the free energy difference for $y$. 
Given that the field introduces drift in the random walk and changes the transition probability to $y$ and the transition probability to $-y$ from the same initial location, this result is fundamentally the same as exchanging the field direction, as shown below. 

By knowing that $g_0(|x-x_{\rm i}|,t)$ is an even function of $\mu $ and defining 
\begin{eqnarray}
G_0(x,x_{\rm i},F ,t)=\exp\left[(\mu /\sigma^2  )(x-x_{\rm i})\right]g_0(|x-x_{\rm i}|,t) ,
\label{eq_A}
\end{eqnarray}
where $\mu$ is an odd function of $F$ as shown in Eq. (\ref{eq:Aapprox}),
we obtain
\begin{eqnarray}
\frac{G_0(x,x_{\rm i},F ,t)}{G_0(x,x_{\rm i},-F ,t)}=\exp \left(\frac{qF(x-x_{\rm i})}{k_{\rm B} T}\right) . 
\label{eq;flA}
\end{eqnarray}
Equation (\ref{eq;flA}) indicates that the ratio for the trajectory distribution by changing the drift (field) direction is equal to the ratio for the stationary distribution expressed in terms of the free energy difference resulting from the applied external electric field. 
In free space without a boundary, exchange of the carrier locations is equivalent to changing the field (or drift) directions. 
Therefore, we cannot distinguish whether $\exp \left[qF(x-x_{\rm i})/(k_{\rm B} T)\right]$ is related to the distribution obtained by exchanging the initial and final positions [Eq. (\ref{eq;fl})] or to the distribution obtained by exchanging drift directions [Eq. (\ref{eq;flA})]. 
In Sec. \ref{sec:refl_c}, we show that the relation given by Eq. (\ref{eq;fld}) and the relation given by Eq. (\ref{eq;flA}) break down when a reflecting boundary is present, whereas Eq. (\ref{eq;fl}) holds irrespective of the reflecting boundary condition in a continuum limit.

Equation (\ref{eq;fl}) indicates that the ratio for the trajectory distribution obtained by exchanging the initial and final positions is equal to the ratio for the stationary distribution expressed in terms of the free energy difference resulting from the applied external electric field. \cite{Crooks_98,Crooks_99,Crooks_00} 
Equation (\ref{eq;fl}) is independent of the form of the waiting-time distribution. 
The waiting-time distribution for transitions to neighboring lattice sites can be arbitrary because  the form of the density of states denoted by $g(E)$ in Eq. (\ref{eq:psitt}) has not been assumed.
The fluctuation relation given by Eq. (\ref{eq;fl}) is independent of the memory kernel in the drift diffusion equation in free space. 
Irrespective of the memory kernel, we proved the Crooks fluctuation theorem, where the ratio between the distribution of trajectories is characterized by the free energy difference between the initial and the final position induced by the applied electric field. \cite{Crooks_98,Crooks_99,Crooks_00} 

Before closing this section, we note that a power-law waiting-time distribution can be obtained when the density of states is given by \cite{Scher_75} 
\begin{eqnarray}
g(E)= \exp \left( - E/E_0 \right) /E_0 .
\label{traped}
\end{eqnarray}
Using Eq. (\ref{releasert}), we obtain the waiting-time distribution function given by Eq. (\ref{eq:psitt}) 
using a dispersive parameter ($\alpha \equiv k_{\rm B}T/E_0$) as \cite{Schnorer_88,Jakobs_93,Seki_03_1,Seki_03_2,Seki_23} 
\begin{eqnarray}
\psi_{\rm t} (t,F) &= 
\int_0^{\infty} d\,E g(E) \gamma_{\rm t} (E,F)  \exp \left( - \gamma_{\rm t} (E,F)  t \right)\label{eq:psit}\\
&=\frac{\alpha \gamma \left( \alpha + 1, \gamma_{\rm rt}(F) t\right)}{\gamma_{\rm rt}(F)^\alpha t^{\alpha+1}} 
\sim \frac{\alpha \Gamma \left( \alpha + 1\right)}{\gamma_{\rm rt}(F)^\alpha t^{\alpha+1}}, 
\label{eq:psit_1}
\end{eqnarray}
where $\gamma (z, p) \equiv \int_0^p e^{-t} t^{z-1} d\,t \mbox{  for } (\mbox{Re} z > 0)$ and $\Gamma (z)$ are the incomplete Gamma function and the Gamma function, respectively. \cite{NIST}
$\alpha <1$ indicates dispersive transport. 
\section{Influence of reflecting boundary condition}
\label{sec:refl_c}

Here, we study the influence of a reflecting boundary placed at $x=0$ for the case of $x_{\rm i}>0$ and $x>0$. 
In this section, we consider a reflecting boundary condition in a continuum limit, where the lattice spacing denoted by $b$ goes to zero. 
In Sec. \ref{sec:refst}, we show the influence of both the finite $b$ and the reflecting boundary by considering the structure factor on a fluctuation relation. 
The solution in the limit of $b\rightarrow 0$ has been already derived as \cite{Seki_23}
\begin{eqnarray}
G_{\rm r}(x,x_{\rm i},t)=\exp\left(\frac{\mu }{\sigma^2  } (x-x_{\rm i})\right) g_{\rm r}(x,x_{\rm i},t) . 
\label{eq:refl0}
\end{eqnarray}
By substituting Eq. (\ref{eq:refl0}) into the reflecting boundary condition given by,  
\begin{eqnarray}
\left. \mu G_{\rm r}(0,x_{\rm i},t) -\frac{\sigma^2  }{2} \frac{\partial}{\partial x}G_{\rm r}(x,x_{\rm i},t) \right|_{x=0}=0 ,
\label{eq:abs_10_1_0}
\end{eqnarray}
we obtain, 
\begin{eqnarray}
\left. \mu  g_{\rm r}(0,x_{\rm i},t) -\sigma^2   \frac{\partial}{\partial x}g_{\rm r}(x,x_{\rm i},t) \right|_{x=0}=0 .
\label{eq:abs_10_1}
\end{eqnarray}
We express $g_{\rm r}(x,x_{\rm i},t)$ as
\begin{equation}
g_{\rm r}(x,x_{\rm i},t)=g_0(|x-x_{\rm i}|,t)+g_0(x+x_{\rm i},t)+\int_0^\infty d z g_0(x+x_{\rm i}+z,t)\zeta(z),  
\label{eq:abs_11}
\end{equation}
and determine $\zeta(z)$ from the boundary condition. 
We can prove that the first two terms on the right-hand side of Eq. (\ref{eq:abs_11}) cancel each other  
in calculating the second term on the left-hand side of Eq. (\ref{eq:abs_10_1}) 
using 
\begin{eqnarray}
\left. \frac{d}{dx}g_0(x+ x_{\rm i},t)\right|_{x=0}&=-\left. \frac{d}{dx}g_0(x- x_{\rm i},t)\right|_{x=0} .
\label{eq:13}
\end{eqnarray}
Eq. (\ref{eq:13}) follows from 
\begin{eqnarray}
\hspace{-1cm}
\lefteqn{ 
\left. \frac{d}{dx}g_0(x\pm x_{\rm i},t)\right|_{x=0}=\frac{1-\hat{\psi}_{\rm t}}{s}\frac{1}{2\pi} \int_{-\infty}^\infty dk\, \frac{(-ik)\exp(\mp ik x_{\rm i})}{1-\hat{\psi}_{\rm t} \left[1+\mu^2/(2\sigma^2  )+(\sigma^2  /2) k^2 \right]} 
}
\label{eq:12_1}\\
&=&\frac{1-\hat{\psi}_{\rm t}}{s}\frac{1}{2\pi} \int_{-\infty}^\infty dk\, \frac{(\mp ik)\exp(- ik x_{\rm i})}{1-\hat{\psi}_{\rm t} \left[1+\mu^2/(2\sigma^2  )+(\sigma^2  /2) k^2 \right]} .
\label{eq:12_2}
\end{eqnarray}

The rest of the first derivatives can be calculated using partial integration as 
\begin{eqnarray}
\frac{d}{dx} \int_0^\infty &dz g_0(x+x_{\rm i}+z,t)\zeta(z)=\int_0^\infty dz \zeta(z)\frac{d}{d z}g_0(x+x_{\rm i}+z,t) 
\label{eq:abs_14_1}\\
&=-g_0(x+x_{\rm i},t)\zeta(0)-\int_0^\infty dz g_0(x+x_{\rm i}+z,t)\frac{d}{d z} \zeta(z). 
\label{eq:abs_14_2}
\end{eqnarray}
When Eq. (\ref{eq:abs_14_2}) is substituted into Eq. (\ref{eq:abs_10_1}), we find 
\begin{eqnarray}
\frac{d}{d z} \zeta(z)=-\frac{\mu }{\sigma^2  } \zeta(z),
\label{eq:15_1}
\end{eqnarray}
where we have $\zeta(0)=-2\mu /\sigma^2  $. 
By solving Eq. (\ref{eq:15_1}),  $\zeta(z)=-(2\mu /\sigma^2  ) \exp\left[-(\mu /\sigma^2  ) z \right]$ is obtained.
Equation (\ref{eq:abs_11}) can be rewritten as,  \cite{Seki_23}
\begin{eqnarray}
g_{\rm r}(x,x_{\rm i},t)=&g_0(|x-x_{\rm i}|,t)+g_0(x+x_{\rm i},t)-
\nonumber \\
&
\frac{2\mu }{\sigma^2  } \int_0^\infty dz g_0(x+x_{\rm i}+z,t)\exp\left(-\frac{\mu }{\sigma^2  } z \right).
\label{eq:abs_16}
\end{eqnarray}

We note that $g_{\rm r}(x,x_{\rm i},t)=g_{\rm r}(x_{\rm i},x,t)$ and find that 
\begin{eqnarray}
\frac{G_{\rm r}(x,x_{\rm i},t)}{G_{\rm r}(x_{\rm i},x,t)}
=\exp \left[(2\mu /\sigma^2  ) (x-x_{\rm i})
\right]=\exp \left(\frac{qF(x-x_{\rm i})}{k_{\rm B} T}\right), 
\label{eq;flg}
\end{eqnarray}
even under the reflecting boundary condition at $x=0$ and when a memory kernel is present in the drift-diffusion equation. 
In Eq. (\ref{eq:refl0}), $g_{\rm r}(x,x_{\rm i},t)$ given by Eq. (\ref{eq:abs_16}) is not 
a function of $x-x_{\rm i}$ alone and Eq. (\ref{eq;flA}) does not hold. 
Similarly, in Eq. (\ref{eq:refl0}), $\hat{g}_{\rm r}(x,x_{\rm i},s)$ expressed by Eq. (\ref{eq:abs_16}) is not an even function of $\mu $. 
Therefore, if $G_{\rm r}(x,x_{\rm i},t)$ is used instead of $G_0(x,x_{\rm i},t)$, then Eq. (\ref{eq;flA}) does not hold. 

\section{Influence of structure factor}
\label{sec:st}

Here, we consider the Green's function given by Eq. (\ref{eq:G0_0d}) with the structure factor given by Eq. (\ref{eq:lambda}), without assuming $kb\ll1$. 
We transform Eq. (\ref{eq:G0_0d}) into a form suitable to study the fluctuation relation.
By introducing $k_2=k-iC$ in Eq. (\ref{eq:lambda}), we obtain 
\begin{eqnarray}
\lambda(k_2+iC)=\Gamma_{\rm p} (F) \exp\left(i k b-C b \right) +\Gamma_{\rm m} (F)\exp\left(-i k b +Cb\right) .
\label{eq:lambdaC}
\end{eqnarray}
We determine $C$ to satisfy $\Gamma_{\rm p} (F) \exp\left(-C b \right) =\Gamma_{\rm m} (F)\exp\left(Cb\right)$ so that the denominator of Eq. (\ref{eq:G0_0d}) becomes an even real function of $k$. 
When $ \int_{-\pi/b}^{\pi/b} dk\, \exp[-ik(x-x_{\rm i})]$ is applied to an even real function of $k$, the result should be an even real function of $x-x_{\rm i}$, which is a desirable form to study the effect of exchanging $x$ and $x_{\rm i}$. 
Using Eq. (\ref{eq:lambdaC}) and Eq. (\ref{eq:15_5_st}), we find that
\begin{eqnarray}
C=\frac{1}{2b} \ln \frac{\Gamma_{\rm p} (F)}{\Gamma_{\rm m} (F)}=qF/(2 k_{\rm B} T).
\label{eq:c}
\end{eqnarray}
Using Eq. (\ref{eq:c}), we can rewrite $\lambda(k)$ as
\begin{eqnarray}
\lambda(k_2+iC)&=\frac{\cos(k_2 b)}{\cosh\left[ q Fb/(2k_{\rm B} T)\right] }.
\label{eq:lambdaC1}
\end{eqnarray}
The Green's function can then be expressed as
\begin{eqnarray}
G_0(x,x_{\rm i},t)=\exp\left[C(x-x_{\rm i})\right]g_0(|x-x_{\rm i}|,t),  
\label{eq_7_1}
\end{eqnarray}
where $g_0(|x-x_{\rm i}|,t)$ can be expressed in the Laplace domain as  
\begin{eqnarray}
\hat{g}_0(|x-x_{\rm i}|,s)&= \frac{1-\hat{\psi}_{\rm t}}{s}\frac{1}{2\pi} \int_{-\pi/b}^{\pi/b} dk\, \frac{\exp[-ik(x-x_{\rm i})]}{1-\hat{\psi}_{\rm t} \cos(k b)/\cosh\left[ q Fb/(2k_{\rm B} T)\right] }.
\label{eq:abs_8_1}
\end{eqnarray}
$\hat{g}_0(x-x_{\rm i},s)$ is an even function of $x-x_{\rm i}$. 
Equations (\ref{eq_7_1}) and (\ref{eq:abs_8_1}) are consistent with the previous results. \cite{Berezhkovskii_08}
Using the Green's function, we derive the fluctuation relation: 
\begin{eqnarray}
\frac{G_0(x,x_{\rm i},t)}{G_0(x_{\rm i},x,t)}=\exp\left[2C(x-x_{\rm i})\right]=\exp \left[(x-x_{\rm i})qF/(k_{\rm B} T)\right]. 
\label{eq;fll}
\end{eqnarray}
Therefore, the assumption $kb\ll1$ is not required to derive the fluctuation relation given by Eq. (\ref{eq;fl}).

Equation (\ref{eq_7_1}) is a function of $x-x_{\rm i}$ alone. 
Therefore, Eq. (\ref{eq;fld}) holds as shown previously. \cite{Berezhkovskii_08,burov_22,Hamid_13}
Similarly, in Eq. (\ref{eq_7_1}), $g_0(|x-x_{\rm i}|,t)$ expressed by Eq. (\ref{eq:abs_8_1}) is an even function of $F$. 
Therefore, Eq. (\ref{eq;flA}) holds. 

\section{Influence of reflecting boundary condition and structure factor}
\label{sec:refst}

Here, we study the influence of a reflecting boundary placed at $x=0$ for the case of $x_{\rm i}>0$ and $x>0$ 
in addition to the structure factor, without assuming $kb\ll1$. 
We assume that the solution under the reflecting boundary condition can be given in the form of Eq. (\ref{eq_7_1}) and introduce 
$g_{\rm r}(x,x_{\rm i},t)$ to satisfy, 
\begin{eqnarray}
G_{\rm r}(x,x_{\rm i},t)&=\exp\left[C(x-x_{\rm i})\right]g_{\rm r}(x,x_{\rm i},t) 
\label{eq:refl0_st0}\\
&=\left( \frac{\Gamma_{\rm p} (F)}{\Gamma_{\rm m} (F)}
\right)^{(x-x_{\rm i})/(2b)} g_{\rm r}(x,x_{\rm i},t) ,
\label{eq:refl0_st}
\end{eqnarray}
where we have $C=qF/(2k_{\rm B} T)$ from Eq. (\ref{eq:c}). 
By substituting Eq. (\ref{eq:refl0_st}) into the reflecting boundary condition given by, \cite{vanKampen}  
\begin{eqnarray}
-\Gamma_{\rm p} (F) G_{\rm r}(0,x_{\rm i},t) +\Gamma_{\rm m} (F) G_{\rm r}(b,x_{\rm i},t) =0 ,
\label{eq:abs_10_1_0_st}
\end{eqnarray}
we obtain, 
\begin{eqnarray}
-\Gamma_{\rm p}^{1/2} (F)   g_{\rm r}(0,x_{\rm i},t) +\Gamma_{\rm m}^{1/2} (F) g_{\rm r}(b,x_{\rm i},t) =0 .
\label{eq:abs_10_1_st}
\end{eqnarray}
Equation (\ref{eq:abs_10_1_0_st}) indicates the vanishing of the probability current between the site $0$ and site $b$. \cite{vanKampen} 
In the limit of $b \rightarrow 0$, 
Eq. (\ref{eq:abs_10_1_0_st}) can be approximated by, 
\begin{eqnarray}
-\Gamma_{\rm p} (F) G_{\rm r}(0,x_{\rm i},t) +\Gamma_{\rm m} (F) 
\left[G_{\rm r}(0,x_{\rm i},t) +\left. b \frac{d}{dx} G_{\rm r}(x,x_{\rm i},t) \right|_{x=0}\right]=0 .
\label{eq:abs_10_1_0_st_ap}
\end{eqnarray}
By using $\Gamma_{\rm p} (F) -\Gamma_{\rm m} (F)  \rightarrow qFb/(2k_{\rm B} T)=\mu/b$, and 
$\Gamma_{\rm m} (F) \rightarrow 1/2$, Eq. (\ref{eq:abs_10_1_0_st_ap}) reduces to 
the reflecting boundary condition in the continuum limit given by 
Eq. (\ref{eq:abs_10_1_0}), where $\sigma^2 = b^2$ is used.  

We express $g_{\rm r}(x,x_{\rm i},t)$ as
\begin{equation}
g_{\rm r}(x,x_{\rm i},t)=g_0(|x-x_{\rm i}|,t)+g_0(x+x_{\rm i},t)+\sum_{j=0}^\infty  g_0(x+x_{\rm i}+b j,t)\zeta(j),  
\label{eq:abs_11_st}
\end{equation}
and determine $\zeta(j)$ from the boundary condition, 
where the Laplace transform of $g_0(|x-x_{\rm i}|,t)$ is given by Eq. (\ref{eq:abs_8_1}). 
We first note,    
\begin{eqnarray}
\hat{g}_0(|b\pm x_{\rm i}|,s)=\frac{1-\hat{\psi}_{\rm t}}{s}\frac{1}{2\pi} \int_{-\pi/b}^{\pi/b} dk\, \frac{\exp(\mp ik b-i k x_{\rm i})}{1-\hat{\psi}_{\rm t} \cos(k b)/\cosh\left[ q Fb/(2k_{\rm B} T)\right] } .
\label{eq:12_1_st}
\end{eqnarray}
We consider the second term on the left-hand side of Eq. (\ref{eq:abs_10_1_st}) using the first two terms on the right-hand side of Eq. (\ref{eq:abs_11_st}). 
Using Eq. (\ref{eq:12_1_st}), 
$2 g_b(x_{\rm i},t)=g_0(|b- x_{\rm i}|,t)+g_0(b+x_{\rm i},t)$ in the second term on the left-hand side of Eq. (\ref{eq:abs_10_1_st}) can be expressed 
in the Laplace domain as,
\begin{eqnarray}
\hat{g}_b(x_{\rm i},s)=
\frac{1-\hat{\psi}_{\rm t}}{s}\frac{1}{2\pi} \int_{-\pi/b}^{\pi/b} dk\, \frac{\cos(kb)\exp(-i k x_{\rm i})}{1-\hat{\psi}_{\rm t} \cos(k b)/\cosh\left[ q Fb/(2k_{\rm B} T)\right] }  .
\label{eq:13_st}
\end{eqnarray}

Using the third term on the right-hand side of Eq. (\ref{eq:abs_11_st}), we find 
\begin{eqnarray}
\sum_{j=0}^\infty g_0(b+x_{\rm i} +bj,t)&\zeta(j)=
\sum_{\ell=1}^\infty g_0(x_{\rm i} +b\ell,t)\zeta(\ell-1) 
\nonumber \\
&= \sum_{\ell=0}^\infty g_0(x_{\rm i} +b\ell,t)\zeta(\ell-1) -g_0(x_{\rm i} ,t)\zeta(-1) .
\label{eq:abs_14_2_st}
\end{eqnarray}
When Eq. (\ref{eq:abs_14_2_st}) is substituted into Eq. (\ref{eq:abs_10_1_st}), we note that the following equation should hold, 
\begin{eqnarray}
\frac{\zeta(j)}{\zeta(j-1)}=\left( \frac{\Gamma_{\rm m} (F)}{\Gamma_{\rm p} (F) }
\right)^{1/2} ,
\label{eq:15_1_st}
\end{eqnarray}
and the rest of the terms should satisfy, 
\begin{eqnarray}
-2\Gamma_{\rm p}^{1/2} (F)  g_0(x_{\rm i},t) +
\Gamma_{\rm m}^{1/2} (F) 
\left[2g_b(x_{\rm i},t) 
-g_0(x_{\rm i} ,t)\zeta(-1) 
\right]=0 ,
\label{eq:15_2_st}
\end{eqnarray}
where $g_b(x_{\rm i},t)$ is given by Eq. (\ref{eq:13_st}). 
Equation (\ref{eq:15_2_st}) yields, 
\begin{eqnarray}
\zeta(-1)=2 \left[\frac{g_b(x_{\rm i},t)}{g_0(x_{\rm i},t)}
-\left( \frac{\Gamma_{\rm p} (F)}{\Gamma_{\rm m} (F) }
\right)^{1/2}
\right] ,
\label{eq:15_3_st}
\end{eqnarray}
and Eq. (\ref{eq:15_1_st}) yields, 
\begin{eqnarray}
\zeta(j)=\left( \frac{\Gamma_{\rm m} (F)}{\Gamma_{\rm p} (F) }
\right)^{(j+1)/2}\zeta(-1) .
\label{eq:15_4_st}
\end{eqnarray}
Using Eq. (\ref{eq:15_5_st}) we obtain, 
\begin{eqnarray}
g_{\rm r}(x,x_{\rm i},t)=&g_0(|x-x_{\rm i}|,t)+g_0(x+x_{\rm i},t)-
\nonumber \\
&
A_b (x_{\rm i}) \sum_{j=0}^\infty g_0(x+x_{\rm i}+b j,t)\exp\left[ -q Fbj/(2k_{\rm B} T)\right]  ,
\label{eq:abs_16_st}
\end{eqnarray}
where $A_b (x_{\rm i})=\exp\left[ -q Fb/(2k_{\rm B} T)\right]\zeta(-1)$ can be expressed as
\begin{eqnarray}
A_b (x_{\rm i})=2\exp\left[ -q Fb/(2k_{\rm B} T)\right]
\left[\exp\left(\frac{qFb}{2k_{\rm B} T} \right)
-\frac{g_b(x_{\rm i},t)}{g_0(x_{\rm i},t)}
\right] .
\label{eq:abs_Ab}
\end{eqnarray}
Because of $\cos(kb)$ in $\hat{g}_b(x_{\rm i},s)$ [Eq. (\ref{eq:13_st})],
we have $g_{\rm r}(x,x_{\rm i},t)\neq g_{\rm r}(x_{\rm i},x,t)$
and find from Eq. (\ref{eq:refl0_st0}) that 
\begin{eqnarray}
\frac{G_{\rm r}(x,x_{\rm i},t)}{G_{\rm r}(x_{\rm i},x,t)}
\neq \exp \left(\frac{qF(x-x_{\rm i})}{k_{\rm B} T}\right), 
\label{eq;flg_st}
\end{eqnarray}
if the limit of $b \rightarrow 0$ is not taken; 
the fluctuation relation given by  
Eq. (\ref{eq;fl}) does not hold when a reflecting boundary condition is imposed by taking into account the structure factor of the lattice. 
One of the merit of studying random walk without taking a continuum limit is that 
we are able to study a nonlinear field driven transport characterized by $qFb/(2k_{\rm B} T)>1$. 
As shown in Sec. \ref{sec:st}, the fluctuation relation holds for open boundaries without a reflecting boundary for $qFb/(2k_{\rm B} T)>1$. 
Here, we show a subtle point in studying a non-linear field driven transport using a continuous time random walk model when a reflecting boundary condition is imposed. 

The solution given by Eqs. (\ref{eq:abs_16_st}) and (\ref{eq:abs_Ab}) reduces to Eq. (\ref{eq:abs_16}) in a continuum limit. 
In the limit of $b \rightarrow 0$, we have $g_b(x_{\rm i},t) \rightarrow g_0(x_{\rm i},t)$, and 
$A_b (x_{\rm i})  \rightarrow qFb/(k_{\rm B} T)$; 
$A_b (x_{\rm i})$ becomes independent of $x_{\rm i}$ and 
the relation given by $g_{\rm r}(x,x_{\rm i},t)=g_{\rm r}(x_{\rm i},x,t)$ is recovered. 
Using $\sum_{j=0}^\infty b f( bj) \rightarrow \int_0^\infty dx f(x)$ for a general function of $f(x)$, and 
$2 \mu/\sigma^2 \rightarrow qF/(k_{\rm B} T)$ [Eqs. (\ref{eq:Aapprox}) and (\ref{eq:Bapprox})], 
we note that Eq. (\ref{eq:abs_16}) is obtained by taking the limit of $b \rightarrow 0$ for Eqs.  (\ref{eq:abs_16_st}) and (\ref{eq:abs_Ab}). 
$g_{\rm r}(x,x_{\rm i},t) \neq g_{\rm r}(x_{\rm i},x,t)$ follows from $\cos(kb)$ in $\hat{g}_b(x_{\rm i},s)$; 
by taking the limit of $b \rightarrow 0$, the fluctuation relation given by Eq. (\ref{eq;fl}) is recovered.  

\section{Conclusion}

Stochastic thermodynamics was originally formulated using Langevin equations, where the equation of motion was used. \cite{Sekimoto_97,Sekimoto_98}
Thermodynamic work can be straightforwardly expressed using the equation of motion in a Langevin equation. 
The ensemble-averaged quantities calculated from a Langevin equation can be equivalently obtained using the Fokker--Planck equation. 
Therefore, the Fokker--Planck equation is also amenable to stochastic thermodynamics. 
The lattice random walks considered here are not directly related to a Langevin equation but are related to Fokker--Planck equations as the lattice spacing approaches zero. 
By considering arbitrariness in defining the work done in stochastic thermodynamics, \cite{Horowitz_07,Seifert_08} which can be introduced only through stationary distributions for lattice random walks, we focus on a fundamental fluctuation relation under nonequilibrium situations expressed using transition probability during an arbitrary time duration.

Using a continuous-time lattice random walk model under a general waiting-time distribution of transitions, we consider noninteracting charge carriers driven by an external electric field. 
 Various non-equilibrium conditions associated with different boundary conditions can be analytically studied by continuous time random walk models.
A fluctuation relation [Eq. (\ref{eq;fl})] is derived by expressing the Green's function for a lattice random walk in a suitable form to study the effect of exchanging the initial and final positions 
for free random walks without boundary conditions. 
In the presence of a reflecting boundary condition, Eq. (\ref{eq;fl}) is derived in a continuum limit, where the lattice spacing denoted by $b$ goes to zero.
When a reflecting boundary condition is imposed without taking the limit of $b \rightarrow 0$, 
Eq. (\ref{eq;fl}) does not hold. 
The result indicates a subtle point in imposing a reflecting boundary condition in continuous time random walk models when particles are driven by a field in a nonlinear 
response regime given by $qFb/(2k_{\rm B} T)>1$.

Equation (\ref{eq;fl}) can be regarded as the Crooks fluctuation theorem in that the ratio between the distribution of trajectories obtained by changing the initial and final positions is characterized by the free energy difference between the initial and final positions induced by the applied field. \cite{Crooks_98,Crooks_99,Crooks_00} 
In free space, the Crooks fluctuation theorem holds irrespective of the lattice structure factor and the form of the waiting-time distribution of transitions. 
In free space, we can also prove Eqs. (\ref{eq;fld}) and (\ref{eq;flA}), which are given by the ratio for the trajectory distribution by changing the drift (field) directions without changing the initial position.
However, Eqs. (\ref{eq;fld}) and (\ref{eq;flA}) break down when a reflecting boundary is present,  
whereas Eq. (\ref{eq;fl}) holds irrespective of the reflecting boundary condition in a continuum limit.  
Therefore, the Crooks fluctuation theorem should be interpreted in terms of exchanging the initial and final positions rather than exchanging the drift (field) directions. 
The left-hand side of Eq. (\ref{eq;fl}) expressed using exchange of the initial and final positions appears also in the detailed balance relation in a closed equilibrium system; \cite{vanKampen} 
the detailed balance originates from microscopic reversibility. \cite{vanKampen,deGroot_62} 
Here, the detailed balance in equilibrium is extended for the nonequilibrium situation driven by an external field 
under various boundary conditions by taking into account a lattice structure factor as well as in the limit of $b \rightarrow 0$. 

The derivation of fluctuation relation relies on the reciprocity (causality) relation of the Lattice Green's function. 
Imposing boundary conditions associated with various non-equilibrium situations can be analytically performed using the Lattice Green's function.
As long as the Lattice Green's function can be introduced, fluctuation relation might be proved for other lattices. 
However, the derivation is not applicable if the Lattice Green's function is not defined. 
If particles are driven by an external field on a lattice, where each site has the fixed random trap energy, the situation is called quenched energy disorder.
For quenched energy disorder, the average with respect to the density of states should, in principle, be used in evaluating the physical quantity, which is the ratio between the normal transition probability and the transition probability obtained by exchanging the initial and final positions; \cite{Krusemann_14,Krusemann_15}
The waiting-time distribution calculated using the density of states can be regarded as pre-averaging of the energetic disorder. 
Studying the fluctuation relation for a lattice random walk under an external electric field with quenched energy disorder without resorting to pre-averaging remains an open problem.



\end{document}